\documentclass[12pt]{article}
\pdfoutput=1
\usepackage[centertags]{amsmath}
\usepackage[square, comma, sort&compress,numbers]{natbib}
\usepackage{array,multirow}
\numberwithin{equation}{section}
\usepackage{amssymb,amsmath,amsthm}
\usepackage{graphicx}
\usepackage{color}
\usepackage[normalem]{ulem}

\usepackage{epsfig}
\usepackage{bbold}

\usepackage{wrapfig}
\usepackage{float}
\usepackage{soul}

\usepackage{tikz,pgf}
\usetikzlibrary{shapes}
\usetikzlibrary{calc}
\usetikzlibrary{decorations.pathmorphing}
\usetikzlibrary{decorations.pathreplacing,shapes.misc}
\usetikzlibrary{positioning}
\usetikzlibrary{arrows}
\usetikzlibrary{decorations.markings}
\usetikzlibrary{shadings}

\usetikzlibrary{intersections}

\def\be{\begin{equation}}
\def\ee{\end{equation}}
\def\ba{\begin{array}}
\def\ea{\end{array}}

\def\dps{\displaystyle}

\def\sd{D^\dagger}

\newcommand{\brst}{\mathsf{\Omega}}

\def\half{\frac{1}{2}}

\def\cN{\mathcal{N}}

\newcommand{\fo}{\theta}
\newcommand{\ao}{\Gamma}
\newcommand{\Dirac}{\widehat{D}}

\usepackage{jheppub}
\usepackage{amsmath,amsfonts,amssymb,amsthm}
\newcommand{\bref}[1]{\textbf{\ref{#1}}}

\newcommand{\dl}[1]{\frac{\dd}{\dd #1}}

\newcommand{\dd}{\partial}
\renewcommand{\d}{\partial}
\newcommand{\gh}[1]{\mathrm{gh}(#1)}
\renewcommand{\st}[2]{\overset{#1}{#2}}
\newcommand{\inner}[2]{\langle #1{,}\,#2\rangle}
\def\cE{\mathcal{E}}
\def\cF{\mathcal{F}}
\def\cG{\mathcal{G}}
\def\cH{\mathcal{H}}

\def\cM{\mathcal{M}}
\def\cN{\mathcal{N}}

\def\cP{\mathcal{P}}

\def\cS{\mathcal{S}}

\usepackage{mathtools}	% for := as \coloeqq
\usepackage{bm}

\newcommand*\dif{\mathop{}\!\mathrm{d}}

\newcommand{\Liealg}{\mathfrak}
\newcommand{\Lorentz}{\Liealg{so}(d-1,1)}
\newcommand{\Howe}{\Liealg{osp}(1|2n+2)}

\newcommand{\fC}{\mathbb{C}}

\makeatletter
\def\@fpheader{\vspace{-.1cm}}
\makeatother

\title{Unified formulation for helicity and continuous spin fermionic fields}

\author[a,b]{Konstantin\ Alkalaev,}

\author[a,c]{Alexander \ Chekmenev,}

\author[a]{Maxim\ Grigoriev}

\affiliation[a]{I.E. Tamm Department of Theoretical Physics, \\P.N. Lebedev Physical
  Institute,\\ Leninsky ave. 53, 119991 Moscow, Russia}
\affiliation[b]{Department of General and Applied Physics, \\
  Moscow Institute of Physics and Technology, \\
  Institutskiy per. 7, Dolgoprudnyi, \\141700 Moscow region, Russia}
\affiliation[c]{Department of Higher Mathematics, \\
  Moscow Institute of Physics and Technology, \\
  Institutskiy per. 7, Dolgoprudnyi, \\141700 Moscow region, Russia}
\emailAdd{alkalaev@lpi.ru}
\emailAdd{chekmenev@phystech.edu}
\emailAdd{grig@lpi.ru}

\abstract{We propose a unified BRST formulation of general massless fermionic fields of arbitrary mixed-symmetry type in $d$-dimensional Minkowski space. Depending on the value of the real parameter the system describes either helicity fields or continuous spin fields. Starting with the unified formulation we derive a number of equivalent descriptions including the triplet formulation, Fang-Fronsdal-Labastida formulation, light-cone formulation and discuss the unfolded formulation.}
  
\preprint{FIAN-TD-2018-14}

\begin{document}

\maketitle
\flushbottom

\section{Introduction and summary}

Continuous spin fields provide an interesting example of field theory systems with an infinite number of physical degrees of freedom \cite{Wigner:1939cj,Bargmann:1948ck} (for recent review see \cite{Bekaert:2017khg}). Group-theoretically, continuous spin particles are unitary representations of Poincar\'e algebra $\Liealg{iso}(d-1,1)$,  induced from unitary representations of the stability subalgebra $\Liealg{iso}(d-2)\subset \Liealg{iso}(d-1,1)$.\footnote{See e.g.~\cite{Bekaert:2006py} for a review of the Poincar\'e representations relevant in the present context.}   Contrary to the standard helicity fields where one induces from finite-dimensional unitary representations of the little algebra $\Liealg{o}(d-2)$, the continuous spin representations correspond to infinite-dimensional $\Liealg{iso}(d-2)$-modules. 

A continuous spin parameter denoted by a real number $\mu$ is an eigenvalue of the squared $\Liealg{iso}(d-2)$ momentum or of the quartic $\Liealg{iso}(d-1,1)$ Casimir operator \cite{Brink:2002zx}. Remarkably, the standard mass parameter associated to the quadratic $\Liealg{iso}(d-1,1)$ Casimir operator is zero, $m=0$, so that the continuous spin fields are massless fields simultaneously characterized by the dimensionful  parameter $\mu$.\footnote{In this respect, the original term "continuous spin" is  somewhat misleading because such systems behave like massive ones. For instance, they can be obtained through a dimensional reduction of the standard higher spin massive systems where the mass $m\to 0$ and the spin $s\to \infty$, while a combination $ms$ is kept finite \cite{Khan:2004nj,Bekaert:2005in}.}

Since the original Wigner's equations for the continuous spin fields were proposed \cite{Wigner:1939cj,Bargmann:1948ck}, several interesting descriptions  were developed both at the level of  equations of motion  \cite{Bekaert:2005in,Bengtsson:2013vra,Najafizadeh:2017tin,Alkalaev:2017hvj,Buchbinder:2018soq} and of the action functional \cite{Schuster:2014hca,Rivelles:2014fsa,Najafizadeh:2015uxa,Metsaev:2016lhs,Zinoviev:2017rnj,Khabarov:2017lth,Metsaev:2017ytk,Metsaev:2018lth,Buchbinder:2018yoo}. A characteristic feature of these formulations is that the space of fields is infinite-dimensional~\cite{Bekaert:2005in} in accord with infinite dimensionality of the respective little group representation.

The relation between the continuous spin and usual helicity fields becomes manifest within the Schuster-Toro formulation \cite{Schuster:2014hca,Metsaev:2016lhs}. In this approach a single continuous spin field is described by an infinite collection of Fronsdal tensors with ranks running from zero to infinity, making it quite similar to the standard interacting higher spin theory~\cite{Fradkin:1987ks,Vasiliev:2003ev} (for a review see e.g. \cite{Bekaert:2005vh}) whose free limit is an infinite tower of the helicity spin fields.\footnote{Note that continuous spin fields themselves can consistently interact with massive higher spin fields, at least in the cubic order~\cite{Metsaev:2017cuz,Bekaert:2017xin}.} In particular, for vanishing continuous spin parameter $\mu=0$  Schuster-Toro system decomposes into an infinite collection of decoupled Fronsdal spin-$s$ fields with $s=0,1,2,...,\infty$. In this regard, the continuous spin field is somewhat similar to the standard Minkowski space massive field which also decomposes into a collection of massless ones in the zero mass limit, see e.g.~\cite{Zinoviev:2001dt}. One can also draw an analogy with  a generic massless higher spin field in AdS that in the flat limit decomposes into a collection of Minkowski space massless fields~\cite{Brink:2000ag}.        

A rather concise unified BRST formulation of the bosonic continuous spin fields that explicitly manifests all of the above features has been proposed recently in~\cite{Alkalaev:2017hvj}. It is based on a constrained system which is a deformation of the one employed in studying mixed symmetry helicity fields~\cite{Alkalaev:2008gi}. This formulation is suitable for analyzing the content of the system through studying its BRST cohomology. In particular, in this way it was shown that with the naive choice of the functional class in the sector of auxiliary variables the system is pure gauge, i.e. there are no degrees of freedom. Nevertheless, it turns out  that with the proper choice of the functional class the system indeed describes proper degrees of freedom~\cite{Alkalaev:2017hvj}.

An additional attractive feature of the unified formulation is that its different reductions reproduce  various other equivalent formulations including the metric-like \cite{Bekaert:2005in,Schuster:2014hca} and frame-like \cite{Zinoviev:2017rnj,Khabarov:2017lth,Ponomarev:2010st} formulations as well as the light-cone formulation \cite{Metsaev:2017cuz}. 

In this paper we propose a fermionic extension of the unified BRST formulation of both the helicity and continuous spin fields.\footnote{For the previous results on higher spin fermions propagating in Minkowski and (A)dS backgrounds in the frameworks of both metric-like and frame-like formulations see e.g.  \cite{Fang:1978wz,Vasiliev:1987tk,Metsaev:1998xg,Alkalaev:2001mx,Alkalaev:2003qv,Buchbinder:2004gp,Buchbinder:2007vq,Moshin:2007jt,Skvortsov:2008vs,Zinoviev:2009vy,Campoleoni:2009gs,Skvortsov:2010nh,Reshetnyak:2012ec,Reshetnyak:2018fvd,Reshetnyak:2018yhz,Najafizadeh:2018cpu}. Other descriptions of fermionic continuous spin fields can be found in \cite{Bekaert:2005in,Najafizadeh:2015uxa,Khabarov:2017lth,Metsaev:2017ytk}.}  Just like in the bosonic case,  the underlying formulation is the constrained system whose constraints belong to a subalgebra of $\Liealg{osp}(1|2n)$ superalgebra in the representation where $\Liealg{osp}(1|2n)$ and $\Liealg{o}(d-1,1)$ form a reductive dual pair in the sense of Howe~\cite{Howe}.

The paper is structured as follows. There are two main parts divided between the helicity and the continuous spin cases. In  Section  \bref{sec:prelim} we describe  $\Liealg{o}(d-1,1)$-$\Liealg{osp}(1|2n)$ bimodule (which is also a Poincar\'e one) on the functions of auxiliary (anti)commuting variables, which serves as a representation space of the constrained system. In Section \bref{sec:short} we formulate one-parameter constraint  system such that  helicity and continuous spin fields correspond to different values of the parameter.  In two main Sections \bref{sec:FHF} and \bref{sec:Fcon} we build the triplet, metric-like, light-cone formulations for respectively helicity  and continuous spin fermionic fields. The analysis of BRST cohomology is performed in Section~\bref{sec:modules}. Appendices \bref{sec:appA} and \bref{Casimir operators section}  consider  various aspects of the space-time and symplectic (super)algebras. 

\section{Algebraic preliminaries}
\label{sec:prelim}
\subsection{Spinor-tensor fields }
\label{sec:spinortensor}

Let us introduce  Grassmann even variables $a^a_I$ and $\bar a^J_b$, where $a,b=0, ..., d-1$, $\;I,J=0, ...,n$ and Grassmann odd variables $\theta^a$ satisfying the canonical commutation relations
\be 
\label{oscillatros}
  [{\bar a}_a^I, a^b_{J}] = \delta^I_J\,\delta_a^b\,,
  \qquad\qquad
  \{\fo^a, \fo^b\} = 2 \eta^{ab}\,,
\ee
where $\eta^{ab}=(-+\cdots+)$ is the Minkowski tensor. These variables generate the associative algebra which is then promoted to the operator algebra of a quantum constrained system.

Consider the linear  space $\cP^d_{n}(a_I)= \cS \otimes \fC[a_I]$, where $\cS$ is the Dirac representation of the Clifford algebra generated by $\fo^a$ and $\fC[a_I]$ is the space of polynomials in $a_I^a$. Elements of $\cP^d_{n}(a_I)$ have the component form
\begin{equation} \label{pol}
\begin{gathered}
  \psi(a) = e_\alpha \psi^\alpha(a)\,, \\
  \psi^\alpha(a)= \sum_{m_I}  \psi^\alpha{}_{a_1\;\ldots\; a_{m_0};\;\ldots\ldots\;;\, b_1 \;\ldots\; b_{m_{n}}} a^{a_1}_0 \ldots a^{a_{m_0}}_0 
  \;\ldots\;
  a^{b_1}_{n} \ldots a^{b_{m_{n}}}_{n}\;,
 \end{gathered}
\end{equation}
where $m_I\equiv (m_0, ..., m_{n})$ are arbitrary non-negative integers, $e_\alpha$ is a basis in $\cS$, and $\alpha = 1,..., 2^{[d/2]}$ is the Dirac spinor index. It is also useful to regard $\cP^d_{n}(a_I)$ as the space of polynomial functions in $a^a_I$ with values in $\cS$.

The associative algebra generated by $a^a_I$, $\bar a^J_b$ and $\fo^a$ 
can be represented on  $\cP^d_{n}(a_I)$ in a natural way if one defines the action of the generators according to 
\be
a_I^a \psi(a)\coloneqq a_I^a \psi(a)\;,
\qquad
\bar a^I_a \psi(a) \coloneqq  \frac{\partial}{\partial a_I^a} \psi(a)\;,
\qquad
\fo^a \psi(a)^\alpha\coloneqq  (\gamma^a)^\alpha{}_\beta \psi^\beta(a)\;,
\ee
where the gamma-matrices $(\gamma^a)^\alpha{}_\beta$ are defined in terms of the basis $e_\alpha$ in $\cS$ as
$\theta^a e_\beta =(\gamma^a)^\alpha{}_\beta e_\alpha$.

\subsection{Lorentz algebra and orthosymplectic superalgebra}

The Lorentz algebra $\Lorentz$ can be embedded as a Lie subalgebra into the above operator algebra by postulating 
\begin{equation} \label{lorgen}
  M_{ab} = a_I{}_a \bar a^{I}_b - a_I{}_b \bar a^{I}_a + \frac14 (\fo_a \fo_b - \fo_b \fo_a)\,.
\end{equation}
This also defines a representation of $\Lorentz$ on $\cP^d_{n}(a_I)$. It follows that the expansion coefficients in (\ref{pol}) transform as Lorentz spinor-tensors.

Simultaneously, the orthosymplectic superalgebra  $\Liealg{osp}(1|2n+2)$ can also be embedded into the operator algebra,  and, hence, is also represented on $\cP^d_{n}(a_I)$. The even and odd basis elements are given respectively by 
\be
\label{SPgenerators}
  T_{IJ}=a_I^a a_{Ja}\,,\quad T_I{}^J=\frac{1}{2}\,(a^a_I \bar a^J_a+\bar a^J_a a^a_I)\,,
  \quad T^{IJ}=\bar a^I_a\bar a^{Ja}\,,
\ee
 and 
\begin{equation} \label{Fgenerators}
  \Upsilon_I = a_I^a \fo_a\;,
  \qquad
  \Upsilon^I = \bar a^I_a \fo^a\;,
\end{equation}
with the graded commutation relations given in Appendix \bref{sec:appA}.  The space $\cP^d_{n}(a_I)$ is now $\Lorentz-\Howe$ bimodule. The two algebras mutually commute forming a reductive dual pair~\cite{Howe}.

\subsection{Poincar\'e algebra} 

The Poincar\'e algebra $\Liealg{iso}(d-1,1)$ can be realized on the same set of auxiliary variables. To this end, we split the original  variables as $a^a_0\equiv x^a,\;a^a_I\equiv a^a_i\,,\,I>0$ with $i=1,...,n$. Then, translations and Lorentz rotations are given by 
\begin{equation} \label{Poincare_generators}
  P_a = \partial_a\;, \quad\qquad 
  M_{ab}
  = x_a \partial_b - x_b \partial_a
  + a_i{}_a \bar a^{i}_b - a_i{}_b \bar a^{i}_a
  + \frac14 (\fo_a \fo_b - \fo_b \fo_a)\,,
\end{equation}
and naturally act in the space $\cP_n^d(x,a)$ of smooth functions in $x^a$ with values in $\cP_n^d(a_i)$. 

We also introduce special notation for some of $\Liealg{sp}(2n+2)$ even basis elements 
\be
\label{newnot}
\begin{gathered}
  \Box \equiv T^{00} = \partial_a \partial^a\,,\qquad
  D^i \equiv T^{0i} = \bar a_i^a\partial_a\,, \qquad
  \sd_i\equiv T_i{}^0 = a_i^a \partial_a\,,\qquad \\
  N_i{}^j\equiv T_i{}^j = a_i^a \bar a_{ja} \qquad i\neq j\,,\qquad
  N_i \equiv T_i{}^i - \frac{d}{2} = a_i^a \bar a_{ia}\,,
\end{gathered}
\ee
and for the odd basis element
\begin{equation} 
\label{Fgendef}
\begin{array}{c}
  \Dirac \equiv \Upsilon^0 = \fo^a \partial_a\,.
\end{array}
\end{equation}
In particular, from the $\Howe$ graded commutation relations we have $\{\Dirac, \Dirac\} = 2 \Box$ meaning  that  the Dirac operator  $\Dirac$  squares to the Klein-Gordon operator  $\Box$.

\section{One-parameter family of constraint systems}
\label{sec:short}

We claim that both helicity and continuous spin fermionic fields can be uniformly described by a one-parameter system of constraints which are (deformed) generators of a subalgebra of $\Howe$. The constraints are imposed on a spinor-tensor field $\psi\in \cP_n^d(x,a)$.

The constraint algebra is generated by the Dirac constraint  
\begin{equation}
\label{Dirac}
  \Dirac \psi = 0\;,
\end{equation}
the gamma-trace conditions 
\be
\label{gammatr}
  \left( \Upsilon^i + \nu^i \ao \right) \psi = 0\,, \qquad \nu^i = \nu \delta^{i1}, \qquad i = 1,\ldots,n\,,\\
\ee
and the spin weight and Young symmetry conditions
\be
\label{youngs}  N_m \psi = s_m \psi\,, \qquad N_m{}^k \psi = 0 \quad (m<k)\,, \qquad m,k = 2,\ldots,n\;.
\ee
Here, $\nu \in \mathbb R$, spin weights $s_m \in \mathbb{N}$,  and $\ao$ is the extra Clifford element satisfying $\{\Gamma,\fo^a\}=0$ and $\ao^2 = 1$.

The additional constraint is implemented in a dual way through the equivalence relation determined by the following gauge transformation law 
\begin{equation} \label{gauge}
  \delta \psi = \left( D^\dag_i + \mu_i \right) \chi^i\,, \;\qquad \mu_i = \mu \delta_{i 1}\,, \qquad i = 1, \ldots, n\,.
\end{equation}
Here, $\mu \in \mathbb R$, and $\chi^i\in \cP_n^d(x,a)$ are the gauge parameters satisfying relations following  from the gauge invariance of the differential/algebraic constraints  \eqref{Dirac}-\eqref{youngs}.

The complete set of constraints also involves 
\begin{equation}
 \Box \psi=0\,, \qquad D^i \psi =0\,, \qquad (T^{ij} + \nu^i \nu^j)\psi=0\;,
\end{equation} 
which are consequences of~\eqref{Dirac} and \eqref{gammatr}.  Indeed, $\Dirac^2 = \Box$, $\{\Dirac, \Upsilon^i + \nu^i \ao\} = 2 D^i$, $\{ \Upsilon^i + \nu^i \ao, \Upsilon^j + \nu^j \ao \} = 2 (T^{ij} + \nu^i \nu^j)$.
In what follows, it is also useful  to split the constraints into differential ones that necessarily involve space-time derivatives $\d_a$ and the algebraic constraints that involve only $a_i^a$ and $\theta^a$ auxiliary variables.

\vspace{2mm}

A few comments are in order. 

\paragraph{$\;\;\bullet$}  At $\mu,\nu = 0$  the system enjoys extra reducibility which can be removed by imposing 
in addition $N_1 \psi = s_1 \psi$ and  $N_1{}^k \psi = 0$, where $k = 2,\ldots,n$ and $s_1 \in \mathbb{N}$.
Then, the resulting constraint system describes fermionic helicity fields (see Section \bref{sec:FHF}). 
For $\mu,\nu \neq 0$ the additional constraints are not consistent with the gauge transformations.

\paragraph{$\;\;\bullet$} The extra Clifford element $\Gamma$ in \eqref{gammatr} is introduced to have a parity-preserving deformation of the Grassmann odd elements of the constraint superalgebra. Definition of $\Gamma$ depends on whether the spacetime dimension $d$ is even or odd. More precisely, for even $d$ the $\Gamma$ can be chosen as the "fifth gamma" $\Gamma \coloneqq \Gamma_{d+1}$, where 
\begin{equation}
  \ao_{d+1}
= \frac{i^{d/2 - 1}}{d!} \sqrt{- \det \eta_{ab}}\, \varepsilon_{a_1 \ldots a_d} \fo^{a_1} \ldots \fo^{a_d}
= i^{d/2 - 1} \fo^0 \fo^1 \ldots \fo^{d - 1}\,,
\end{equation}
that is $\Gamma$ can be realized in terms of the original Clifford algebra \eqref{oscillatros} and its module. In odd $d$ it is not the case and $\Gamma$ extends the original Clifford algebra to  $\{\fo^A, \fo^B\} = 2 \eta^{AB}$, where $A = (a, d)$, and $\eta^{dd} = 1$, and $\fo^{d} \coloneqq\Gamma$. In this case, the spinor representation also gets extended, and, hence, the spectrum of fields is duplicated. However, the extended Clifford algebra is even dimensional, and, therefore, there is a new "fifth gamma" $\widetilde\Gamma = i\Gamma_{d+1} \Gamma$ that can be used to project out a half of the spinor components via the standard $P_{\pm} = \half(1\pm \widetilde\Gamma)$.  For simplicity, throughout the paper we explicitly treat the case of even $d$ unless otherwise indicated.
   
\paragraph{$\;\;\bullet$} The values of the quadratic and quartic Casimir operator of the Poincar\'e algebra  evaluated on the subspace \eqref{Dirac}--\eqref{gauge} are given by (see Appendix~\bref{Casimir operators section})
\begin{equation}
\label{casimir4}
C_2 \Big( \mathfrak{iso}(d-1,1) \Big) \psi = 0\;,
\qquad 
C_4 \Big( \mathfrak{iso}(d-1,1) \Big) \psi =  - \mu^2\nu^2  \psi\;.
\end{equation}
At $\mu,\nu \neq 0$ we find out that the above constraint system describes massless fields  characterized by the continuous spin parameter $\mu\nu$. Thus, we indeed have one-parameter constraint system.\footnote{Note that both helicity and continuous spin systems can be obtained by quantizing the spinning particle models \cite{Lyakhovich:1996we}. The resulting field theory is formulated in terms of field strengths rather than gauge fields.} At $\mu, \nu = 0$ the constraint system describes a collection of higher spin massless fields with arbitrary  half-integer helicity. In this case, the eigenvalue in \eqref{casimir4} is zero what exactly matches vanishing eigenvalue of the quartic Casimir operator in the helicity case.

\vspace{3mm}

For general parameters $\mu, \nu$ we fix the functional class in $a_i^a$ to be that of formal power series in $a_i^a$  such that a decomposition of a given element $\psi$ with respect to traces, i.e.
\begin{equation}
 \psi = \psi_0+T_{ij}\psi_1^{ij}+T_{ij}T_{kl}\psi_2^{ijkl}+\ldots\,, \qquad T^{mn}\psi_k^{ij\ldots }=0
 \end{equation} 
 is such that all coefficients are polynomials of bounded order (that means that for a given $\psi$ there exists $N \in \mathbb N$ such that any $\psi_{k}^{ij {\ldots}}$ is of order not exceeding $N$). This functional class was introduced in~\cite{Alkalaev:2017hvj} in the context of bosonic continuous spin fields. Note that in contrast to
 \cite{Alkalaev:2017hvj} now we are concerned with series with coefficients in $\cS$. Equivalently, one can characterize the functional class using the gamma-trace  decomposition
\begin{equation} 
\label{continuous_gamma-trace_decomposition}
\psi = \sum\limits_{k=0}^{\infty} \Upsilon_{i_1} \ldots \Upsilon_{i_k} \psi_{(k)}^{i_1, \ldots, i_k}\;, \qquad i_1 < i_2 < \ldots < i_k\;, \qquad \Upsilon^i \psi_{(k)}^{^{\ldots}} = 0\;,
 \end{equation}
where all coefficients are also required to be polynomials in $a_i^a$ of finite order.

\section{Fermionic helicity fields}
\label{sec:FHF}

In this section, we explicitly study the constraint system \eqref{Dirac}-\eqref{gauge} in the helicity case, $\mu,\nu = 0$. As we noted before, the constraints can be augmented by adding more algebraic conditions so that the resulting system describes a single massless half-integer spin field.   

The augmented system still contains Dirac equation
\begin{equation}
  \Dirac \psi = 0\;,
\end{equation}
while the complete set of algebraic constraints now reads as
\begin{equation}
\label{helcon}
\Upsilon^i \psi = 0\;,
\qquad
N_i \psi = s_i \psi\;,
\qquad
N_i{}^j \psi = 0 \quad (i < j)\;,
  \qquad i,j = 1,\ldots,n\;.
\end{equation}
The spin weight conditions imposed on each type of  auxiliary variables constrain functions $\psi$ to be homogeneous polynomials in $a_i$. The Young symmetry and gamma-tracelessness conditions are the standard irreducibility  conditions for the $\Liealg{o}(d-1,1)$-representation of spin  $s_1+\frac{1}{2} \geq s_2 +\frac{1}{2}\geq ... \geq s_n+\frac{1}{2}$, where $s_i \in \mathbb{N}$.

The gauge transformations read
\begin{equation}
  \delta \psi = D^\dag_i \chi^i, \qquad i = 1, \ldots, n\;,
\end{equation}
where the gauge parameters $\chi^i$ satisfy the same constraints as the fields $\psi$ except for the spin weight and Young symmetry constraints which are replaced by
\begin{gather}
\label{helcon2}
  N_i \chi^j = (s_i - \delta_i^j) \chi^j,\qquad
  N_i{}^j \chi^k = - \delta^k_i \chi^j \quad (i < j)\,.
\end{gather}
Note that the Klein-Gordon operator, the divergence and the trace conditions are imposed by virtue of $\Dirac^2 = \Box$, $\;\{\Dirac, \Upsilon^i\} = 2 D^i$, $\;\{ \Upsilon^i, \Upsilon^j \} = 2 T^{ij}$.

\subsection{Simplest BRST formulation}

Let us introduce the anticommuting ghost variables $b_i$ of ghost number $\gh{b_i} = -1$. The gauge symmetries can be realized via the BRST operator
\begin{equation} \label{helicity_BRST}
  Q = D^\dag_i \frac{\partial}{\partial b_i}\;,
\end{equation}
which acts on the space of functions $\Psi(x,a|b)$ regarded as functions in $x^a$ taking values in $\cP_n^d(a) \otimes \fC[b_i]$, where $\fC[b_i]$ is the Grassmann algebra generated by $b_1, ..., b_{n}$.

Homogeneous components of $\Psi$ in $b_i$ carry definite ghost degree and are introduced according to
\begin{equation}
  \Psi = \sum_{k=0}^{n} \Psi^{(-k)}\;, \qquad \gh{\Psi^{(-k)}} = -k\;.
\end{equation}
The spinor-tensor field $\psi$ above is identified with the ghost number $0$ component $\Psi^{(0)},$, the gauge parameters are $\Psi^{(-1)}$ component, and the order $k$ reducibility parameters  are the ghost degree $-k$ components.

The function $\Psi$ is subject to the BRST invariant extension of the constraints \eqref{helcon},
\begin{equation}
\label{const-DUN}
  \Dirac \Psi = 0\;,\qquad
  \Upsilon^i \Psi = 0\;,\qquad
  \cN_i \Psi = s_i \Psi\;,\qquad
   \cN_i{}^j \Psi = 0 \quad (i < j)\;,
\end{equation}
where
\begin{equation}
  \cN_i{}^j = N_i{}^j + b_i \frac{\partial}{\partial b_j}\;,\qquad
  \cN_i = N_i + b_i \frac{\partial}{\partial b_i}\;.
\end{equation}
The component form of these constraints reproduce that for fields and gauge parameters \eqref{helcon} and \eqref{helcon2}.

Starting with the above BRST formulation one can systematically rederive unfolded formulation~\cite{Vasiliev:1987tk,Skvortsov:2008vs} of mixed-symmetry  fermionic helicity fields in Minkowski space. Indeed, according to~\cite{Barnich:2004cr} the set of fields of the unfolded formulation is given by cohomology of $Q$ evaluated in the subspace~\eqref{const-DUN}. Strictly speaking in so doing one should also replace $x^a$ with formal coordinate $y^a$ and consider elements that are formal series in $y^a$. Moreover, the nilpotent differential determining the unfolded equations and gauge symmetries is just the differential induced by $dx^a(\dl{x^a}-\dl{y^a})+Q$ in the cohomology of the second term. Note that the first term has an interpretation of the flat connection of the Poincar\'e algebra.
The procedure is a straightforward generalization of the derivation~\cite{Alkalaev:2008gi} of the unfolded formulation for general bosonic helicity fields.

\subsection{Extended triplet formulation}

Let us impose all the differential constraints via BRST operator, while all the algebraic constraints or, more precisely, their appropriate BRST invariant extensions we impose directly in the representation space. In the case of integer spin fields,  this reproduces~\cite{Barnich:2004cr,Alkalaev:2008gi} the triplet formulation
discussed previously in~\cite{Bengtsson:1986ys,Francia:2002pt,Sagnotti:2003qa,Fotopoulos:2008ka,Agugliaro:2016ngl,Sorokin:2018djm}. 
As we are going to see for half-integer spin fields this gives the extended description from which the familiar triplet formulation of~\cite{Francia:2002pt,Sagnotti:2003qa} can be obtained by eliminating auxiliary fields and solving constraints.

The extended triplet BRST operator for fermionic helicity fields is given by
\begin{equation} 
\label{helicity_BRST_full}
  \brst = \alpha \Dirac + c_0 \Box + c_i D^i + D^\dag_i \frac{\partial}{\partial b_i} - \alpha \alpha \frac{\partial}{\partial c_0} - c_i \frac{\partial}{\partial b_i} \frac{\partial}{\partial c_0}\;,
\end{equation}
where in addition to the ghost variables $b_i$ we introduced a new anticommuting ghost variables $c_0,c_i$, $i = 1, ..., n$ and commuting ghost variable $\alpha$ with ghosts numbers $\gh{c_{0,i}} = \gh{\alpha} = 1$. As $\alpha$ is a commuting variable there is an ambiguity in the functional class to work with. We choose functions $\Psi(x,a|\alpha,c_0,c,b)$ to be polynomials in $\alpha$.

BRST operator \eqref{helicity_BRST_full} is defined on the subspace singled out by the following BRST-invariant extended constraints
\begin{equation} 
\label{helicity_algebraic_constraints_BRST_extended}
  \widetilde N_i \Psi = s_i \Psi\;, \qquad
  \widetilde N_i{}^j \Psi = 0 \quad (i < j)\;, \qquad
  \widetilde \Upsilon^i \Psi = 0\;,
\end{equation}
where
\begin{equation}
  \widetilde N_i{}^j = N_i{}^j + b_i \frac{\partial}{\partial b_j} + c_i \frac{\partial}{\partial c_j}\;,\qquad
  \widetilde N_i = \widetilde N_i{}^i\;,\qquad
  \widetilde \Upsilon^i = \Upsilon^i - 2 \alpha \frac{\partial}{\partial c_i} + \frac{\partial}{\partial \alpha} \frac{\partial}{\partial b_i}\;.
\end{equation}
Note that the BRST operator \eqref{helicity_BRST_full} is well-defined and is nilpotent on the entire representation space and not only on the subspace \eqref{helicity_algebraic_constraints_BRST_extended}.

\subsection{Homological reduction and the triplet formulation}
\label{sec:redu}

The triplet formulation can be used as a starting point to obtain various other dynamically equivalent formulations including the metric-like formulation and the light-cone formulation. In so doing it is convenient to employ the method of homological reduction developed in \cite{Barnich:2004cr} (see also~\cite{Barnich:2006pc}) and applied earlier to  bosonic mixed-symmetry fields in Minkowski space \cite{Alkalaev:2008gi,Alkalaev:2017hvj} in a similar framework.  

Let us briefly recap the main ingredients of the homological reduction method. Suppose we have a linear gauge theory $(\cH, \Omega)$ defined in terms of the BRST operator $\Omega$ acting on the representation space $\cH$ graded by the ghost number. Let $\cH$ be split into three subspaces: $\cH = \cE\oplus \cF \oplus \cG$ in such a way 
that a linear operator $\st{\cG\cF}{\Omega}: \cF\to \cG$ defined by $\st{\cG\cF}{\Omega}f= \Pi_\cG\Omega f$ for $f\in \cF$ is invertible. It turns out that all the fields  associated with $\cF$ and $\cG$ are generalized auxiliary fields that are  usual auxiliary fields and Stueckelberg fields as well as the associated ghosts and antifields. Generalized auxiliary fields can be eliminated, resulting in an equivalent formulation $(\cE, \widetilde \Omega)$ of the same theory. The reduced BRST operator can be expressed explicitly as
\begin{equation}
 \widetilde\Omega
  =\st{\cE\cE}{\Omega}
  - \st{\cE\cF}{\Omega}(\st{\cG\cF}{\Omega})^{-1}
  \st{\cG\cE}{\Omega}\,,
\end{equation}
where $\st{\cE\cE}{\Omega}$ and $\st{\cG\cE}{\Omega}$ are the respective components of $\Omega$.

In applications, a triple decomposition of $\cH$ is often determined by a certain piece of $\Omega$. More specifically, suppose that $\cH$ admits an additional  grading 
\be
\cH = \bigoplus_{-N}^\infty \cH_i\,, 
\qquad\quad \text{$N$ is a finite integer}\;,
\ee
such that $\Omega$ decomposes into homogeneous components as follows
\be
\Omega =\Omega_{-1}+ \Omega_0+ \Omega_{1}+ \ldots\;. 
\ee

Then, the lowest grade part  $\Omega_{-1}$ of the BRST operator is nilpotent and defines the triple decomposition according to
\be
{\cal E}\oplus {\cal G} = {\rm Ker}\, \Omega_{-1}\;,
\qquad
{\cal G} = {\rm Im}\, \Omega_{-1}\;,
\qquad
{\cal E} \simeq \frac{{\rm Ker}\, \Omega_{-1}}{{\rm Im}\, \Omega_{-1}}\equiv H(\Omega_{-1})\;.
\ee
Note that the subspaces $\cG\subset  \cH$ and $\cG\oplus \cE \subset \cH$ are defined by $\Omega_{-1}$ unambiguously while the embedding of $\cF$ and $\cE$ into $\cH$ is defined up to an ambiguity. The reduced operator $\widetilde \Omega$ can be explicitly expressed~\cite{Barnich:2004cr} in terms of the inverse of $\st{\cG\cF}{\Omega}_{-1}$.

Typically (though not always) one is interested in local gauge field theories in which case one requires that generalized auxiliary fields can be eliminated algebraically. In our case, it means that additional gradings give rise to $\Omega_{-1}$ that do not involve $x$-differential pieces of the triplet BRST operator. 

\paragraph{Triplet formulation.} 
The triplet and metric-like formulations can be obtained from the extended triplet formulation through the homological reduction by taking as additional degree the homogeneity in $c_0$. The BRST operator \eqref{helicity_BRST_full} then decomposes as $\brst = \brst_{-1} + \brst_0 + \brst_1$ with
\begin{equation} \label{brst_c_0_grading_decomposition}
  \brst_{-1} = - \left( \alpha \alpha + c_i \frac{\partial}{\partial b_i} \right) \frac{\partial}{\partial c_0}\;, \qquad
  \brst_0 = \alpha \Dirac + c_i D^i + D^\dag_i \frac{\partial}{\partial b_i}\;, \qquad
  \brst_1 = c_0 \Box\;,
\end{equation}
and, therefore,  we can reduce the theory to the cohomology $H(\brst_{-1})$.

Because $\Omega_{-1}$ is algebraic it is enough to compute cohomology in the space of $x$-independent elements. Let us decompose a generic element as $\Phi=\phi_0 + c_0 \phi_1$, where $\phi_{0,1}=\phi_{0,1}(a,b,c,\alpha)$. The cocycle and the coboundary condition take the form
\begin{equation} 
Z_+ \phi_1 = 0\,,\qquad \phi_0 \sim \phi_0+Z_+\lambda\,,\qquad Z_+\equiv \alpha\alpha+c_i\dl{b_i}\;.
\end{equation}
It is easy to check that in the space of polynomials in $\alpha$ the first equation implies $\phi_1=0$ while $\phi_0$ can be assumed to be at most linear in $\alpha$ thanks to the second equation. Moreover, each equivalence class has a unique representative which is at most linear in $\alpha$. 

To summarize, the cohomology of $\Omega_{-1}$ is concentrated in degree $0$ and can be realized as a subspace $\cE$ of $c_0$-independent elements which are at most linear in $\alpha$. The space of fields with values in this subspace is equipped with the induced BRST operator $\widetilde \Omega$ which in this case is simply $\Omega_0$ defined on the equivalence classes. In terms of representatives which are at most linear in $\alpha$ it is given by 
\begin{equation}
 \widetilde\Omega (\phi_0+\alpha \phi_1)=\alpha \Dirac \phi_0 - c_i\dl{b_i}\Dirac \phi_1 + (c_i D^i + D^\dag_i \frac{\partial}{\partial b_i})(\phi_0+\alpha \phi_1)\,.
\end{equation} 
The  second term  arises from $\alpha^2\Dirac \phi_1$ which does not belong to $\cE$, and, hence, one needs to pick another representative of the same equivalence class.

To see what are the equations of motion encoded in $\widetilde\Omega$, let us restrict ourselves to totally symmetric fields. The field of ghost degree $0$ is then given by
\begin{equation}
 \Phi=\psi+\alpha b \chi +cb\lambda\;,
\end{equation} 
where components $\psi, \chi, \lambda$ form the fermionic triplet \cite{Francia:2002pt}. The equations of motion $\widetilde\Omega \Phi=0$ take the form of the fermionic triplet equations \begin{equation}
\label{FS}
 \Dirac \psi + D^\dagger \chi=0\,, \qquad \Dirac \chi+ D\psi=D^\dagger \lambda\,, \qquad \Dirac \lambda+D\chi=0\,,
\end{equation} 
familiar from~\cite{Francia:2002pt,Sagnotti:2003qa}. The gauge transformations $\delta_\Xi \Phi = \widetilde\Omega \Xi$, where $\Xi = b \epsilon$ is the ghost number $-1$ field, take the form
\begin{equation}
  \delta \psi = D^\dag \epsilon\,,\qquad
  \delta \chi = - \Dirac \epsilon\,,\qquad
  \delta \lambda = D \epsilon\,.
\end{equation}
The above equations and transformations can be supplemented by algebraic  constraints \eqref{helicity_algebraic_constraints_BRST_extended} which are well-defined in $\cE$. In the case of totally symmetric fields they are given by the spin weight conditions
\be
N \psi = s \psi\;,\qquad N\chi = (s-1)\chi\;, \qquad N \lambda  = (s-2) \lambda\;,\qquad
N \epsilon = (s - 1) \epsilon\;,
\ee
along with the trace conditions
\be
\Upsilon\Upsilon\Upsilon \psi =0\;, \qquad \Upsilon\Upsilon \chi =0\;,\qquad
\Upsilon \lambda = 0\;,\qquad
\Upsilon \epsilon = 0\;.
\ee  
Moreover, it follows that $\chi=-\Upsilon \psi$ so that the first equation in~\eqref{FS} gives the Fang-Fronsdal equations~\cite{Fang:1978wz}
\begin{equation}
 \Dirac \psi-D^\dagger \Upsilon \psi=0\,.
\end{equation} 

Let us now consider the case of mixed-symmetry fields.  Elements of $\cE$  of ghost numbers $0$ and $-1$ are given respectively by
\begin{multline}
\Phi = \psi + \sum_{k=1}^{n} c_{i_1} \ldots c_{i_{k}} b_{j_1} \ldots b_{j_k} \lambda^{i_1 \ldots i_{k} | j_1 \ldots j_k} 
+ \\ + \alpha \sum_{k=1}^{n} c_{i_1} \ldots c_{i_{k-1}} b_{j_1} \ldots b_{j_k} \chi^{i_1 \ldots i_{k-1} | j_1 \ldots j_k}\;,
\end{multline}
\begin{multline}
\Xi = b_j \epsilon^j + \sum_{k=2}^{n} c_{i_1} \ldots c_{i_{k-1}} b_{j_1} \ldots b_{j_k} \epsilon^{i_1 \ldots i_{k-1} | j_1 \ldots j_k}
+\\ +
\alpha \sum_{k=2}^{n} c_{i_1} \ldots c_{i_{k-2}} b_{j_1} \ldots b_{j_k} \xi^{i_1 \ldots i_{k-2} | j_1 \ldots j_k}\;.
\end{multline}
From the algebraic constraints \eqref{helicity_algebraic_constraints_BRST_extended} it follows that the lowest components $\psi$ and $\epsilon^i$ satisfy the triple trace conditions 
\begin{equation}
\label{irrcon0}
  \Upsilon^{(i} \Upsilon^{j} \Upsilon^{k)} \psi = 0\,,\qquad
  \Upsilon^{(i} \epsilon^{j)} = 0\,,
\end{equation}
as well as the Young symmetry and spin conditions
\be
\label{irrcon}
N_i{}^j \psi = 0 \;\;  (i<j)\,,
\;\;\;
N_i \psi = s_i \psi\,,
\;\;\;
N_i{}^j \epsilon^k = - \delta^k_i \epsilon^j \;\; (i < j)\,,
\;\;\;
N_i \epsilon^j = (s_i - \delta_i^j) \epsilon^j\,.
\ee
One concludes that $\psi$ and $\epsilon^i$ are precisely the original Fang-Fronsdal-Labastida spinor-tensor fields and their associated gauge parameters \cite{Fang:1978wz,Labastida:1986zb,Labastida:1989kw}.

To derive the metric-like equations we only need the analog of the first equation in \eqref{FS} which reads as
\begin{equation}
  \Dirac \psi + D^\dag_i \chi^{|i} = 0\,.
\end{equation}
The BRST-extended gamma-trace conditions \eqref{helicity_algebraic_constraints_BRST_extended} imply  $\chi^{|i} = - \Upsilon^i \psi$ thereby giving the reduced equations of motion
\begin{equation} 
\label{helicity_metric-like}
\left( \Dirac - D^\dag_i \Upsilon^i \right) \psi= 0\,.
\end{equation}
This is the Fang-Fronsdal-Labastida equations for mixed-symmetry fermionic helicity fields \cite{Fang:1978wz,Labastida:1986zb,Campoleoni:2009gs}. Note that, just like the standard Dirac equation,  the Fang-Fronsdal-Labastida equation can be squared, resulting in
\begin{equation}
  \left( \Box - D^\dag_i D^i + \frac12 D^\dag_i D^\dag_j T^{ij} \right) \psi = 0\,,
\end{equation}
which is the Labastida equations for mixed-symmetry bosonic helicity fields \cite{Labastida:1989kw}. Here we made use 
of $\Dirac D^\dag_i \Upsilon^i \psi = \Box \psi$ which is the result of acting by $\Dirac$ on \eqref{helicity_metric-like}.

By construction, the reduced equations \eqref{helicity_metric-like} are invariant with respect to the gauge transformations
\begin{equation}
\delta \psi = D^\dag_i \epsilon^i\,,
\end{equation}
where the gauge fields and parameters satisfy the algebraic conditions \eqref{irrcon0}--\eqref{irrcon}.

\subsection{Triplet Lagrangian}

The triplet BRST formulation in terms of $\cE$-valued fields and BRST operator $\widetilde\Omega$ can be made Lagrangian by observing that $\cE$ is equipped with a  natural non-degenerate inner product. Let us consider the operator algebra generated by $a_i^a,c_i,b_i,$ as well as their canonically conjugated variables denoted by $\bar a_a^i,\bar c^i,\bar b^i$, and consider the following involution $\dag$\;:
\begin{equation}
\begin{gathered}
	(a_i^a)^\dag = \bar a^i_a\,,\;\;
  (\bar a^i_a)^\dag = a_i^a\,,\;\;
  (c_i)^\dag = - \bar b^i\,,\;\;
  (\bar b^i)^\dag = - c_i\,,\;\;
  (b_i)^\dag = - \bar c^i\,,\;\;
  (\bar c^i)^\dag = - b_i\,,\\
  (A B)^\dag = B^\dag A^\dag\,,
\end{gathered}
\end{equation}
where $A, B$ are generic elements of the algebra. Note that this involution is compatible with the  notations for $D^i, D^\dagger_i$ employed before. Consider the following Fock space generated by these operators from the following vacuum
\begin{equation}
 \bar a_a^i,\bar b^i,\bar c^i|0\rangle=0\,.
 \end{equation} 
Its elements can be identified with the polynomials in $a_i^a,b_i,c_i$ while $\bar a_a^i,\bar c^i,\bar b^i$
are represented by $\dl{a^a_i}$, $\dl{c_i}$, $\dl{b_i}$. 

The above involution uniquely determines the inner product on the Fock space which makes $\dagger$ into conjugation. Tensoring this Fock space with $\cS$ (see Section \ref{sec:spinortensor}) and equipping $\cS$ with an inner product such that $(\theta^a)^\dagger=-\theta^a$ we end up with the space equipped with an inner product $\inner{}{}^\prime$. Finally, the formal inner product on $\cE$ is taken to be
\begin{equation}
\label{inner}
	\langle \phi, \psi \rangle \equiv \langle \phi_0 + \alpha \phi_1, \psi_0 + \alpha \psi_1 \rangle
\coloneqq \int \dif^d x \big( \langle \phi_0, \psi_1 \rangle^\prime
        + \langle \phi_1, \psi_0 \rangle^\prime \big)\,.
\end{equation}
It is straightforward to see that $\widetilde \Omega$ is formally symmetric with respect to the inner product \eqref{inner}. Indeed, the only nontrivial part is to show that $\alpha^\dag = \alpha$ which is clear from the following explicit expressions
\begin{equation}
\begin{gathered}
  \langle \alpha \phi, \psi \rangle
= \langle \alpha \phi_0 - c_i \frac{\partial}{\partial b_i} \phi_1, \psi_0 + \alpha \psi_1 \rangle
= \int \dif^d x \big( \langle \phi_0, \psi_0 \rangle^\prime - \langle c_i \frac{\partial}{\partial b_i} \phi_1, \psi_1 \rangle^\prime \big)\,,\\
  \langle \phi, \alpha \psi \rangle
= \langle \phi_0 + \alpha \phi_1, \alpha \psi_0 - c_i \frac{\partial}{\partial b_i} \psi_1 \rangle
= \int \dif^d x \big( \langle \phi_0, \psi_0 \rangle^\prime - \langle \phi_1, c_i \frac{\partial}{\partial b_i} \psi_1 \rangle^\prime \big)\,,
\end{gathered}
\end{equation}
and $(c_i \frac{\partial}{\partial b_i})^\dagger=(c_i \frac{\partial}{\partial b_i})$.

The equations of motion $\widetilde\Omega \Psi^{(0)} = 0$ for $\Psi^{(0)}$ taking values in $\cE$ follows from the action
\begin{equation}
  S = \frac12 \langle \Psi^{(0)}, \widetilde\Omega \Psi^{(0)} \rangle\;,
\end{equation}
where $\Psi^{(0)}$ is the ghost number 0 field. Moreover, the above action is invariant under the gauge transformations~$\delta \Psi^{(0)}=\widetilde\Omega \Psi^{(-1)}$. It can be supplemented by the off-shell constraints \eqref{helicity_algebraic_constraints_BRST_extended} in which case it is equivalent to the Fang-Fronsdal-Labastida action.

\subsection{Light-cone formulation}
\label{sec:light_helicity}

Starting from the triplet BRST operator \eqref{helicity_BRST_full} and eliminating  unphysical degrees of freedom by means of the homological reduction we arrive at the standard light-cone formulation for half-integer spin massless fields (see, e.g., \cite{Siegel:1988yz}).  This is done by employing the approach developed in~\cite{Barnich:2005ga,Alkalaev:2008gi} (see also~\cite{Aisaka:2004ga} for an earlier important contribution).

As usual, the light-cone coordinates are introduced as  $(x^+,x^-,x^m)$, $m=1,...,d-2$. The light-cone description of the Clifford elements is more tricky. Let us represent $\fo^+$ and $\fo^-$ on the Grassmann algebra $\mathbb C[\fo^+]$ generated by $\fo^+$ as $\fo^+$ and $2 \frac{\partial}{\partial \fo^+}$, respectively. Then, consider the representation of the Clifford algebra with generators $\fo^+, \fo^-, \fo^m$ as a tensor product of $\mathbb C[\fo^+]$ and irreducible representation of the Clifford algebra generated by $\fo^m$. In this way, we realize the representation of the Clifford algebra with generators $\fo^+, \fo^-, \fo^m$ as polynomials in $\fo^+$ with coefficients in $\Liealg{o}(d-2)$ spinors $\psi^{\hat \alpha}$, where the Dirac spinor index $\hat \alpha = 1,..., 2^{[d/2]-1}$. In other words, the light-cone spinor is half of the original spinor.  

To do the light-cone reduction of the fermionic triplet formulation we introduce the grading (this is a generalization of the one employed in~\cite{Aisaka:2004ga,Barnich:2005ga,Alkalaev:2008gi})
\begin{equation} \label{helicity_light-cone_grading}
\begin{gathered}
  \deg a_i^\pm = \pm 2\;,	\quad
  \deg c_i = 1\;,					\quad
  \deg b_i = -1\;,				\quad
  \deg \fo^+ = 2\;,				\quad
  \deg \alpha = 1\;,\\
  \deg a^m = 0, \qquad \deg c_0 = 0\;.
\end{gathered}
\end{equation}

The operator \eqref{helicity_BRST_full} decomposes into the homogeneous degree components as $\brst = \brst_{-1} + \brst_0 + \brst_1 + \brst_2 + \brst_3$, where
\begin{equation}
\label{lightconebrst}
\begin{gathered}
  \brst_{-1} = p^+ \left( 2 \alpha \frac{\partial}{\partial \fo^+} + c_i \frac{\partial}{\partial a_i^+} + a_i^- \frac{\partial}{\partial b_i} \right)\,,\qquad
  \brst_0 = c_0 \Box\,,\\
  \brst_1 = \alpha \fo^m p_m + c_i p^m \frac{\partial}{\partial a_i^m} + a_i^m p_m \frac{\partial}{\partial b_i}\,,\qquad
  \brst_2 = - \left( \alpha \alpha + c_i \frac{\partial}{\partial b_i} \right) \frac{\partial}{\partial c_0}\,,\\
  \brst_3 = p^- \left( \alpha \fo^+ + c_i \frac{\partial}{\partial a_i^-} + a_i^+ \frac{\partial}{\partial b_i} \right)\,.
\end{gathered}
\end{equation}
We assume $p^+ \ne 0$. Then, we observe that $\brst_{-1}$ is just the de Rham differential on the superspace so that the only non-vanishing cohomology  $H(\brst_{-1})$ is in degree $0$ and can be identified with the subspace $\cE$ of elements $\psi^{\hat \alpha} = \psi^{\hat \alpha}(x|a^m,c_0)$ depending on the spacetime coordinates, transverse auxiliary variables, and ghost $c_0$ taking values in the representation of the Clifford algebra generated by transverse $\fo^m$ (in what follows we omit the spinor index).

The cohomology of $\brst_{-1}$ is concentrated in one degree so that the reduced operator $\widetilde \brst$ is given by (see e.g.~\cite{Barnich:2005ga})
\begin{equation}
  \widetilde \brst = c_0 \Box\;.
\end{equation}
The reduced form of the light-cone algebraic constraint
\eqref{helicity_algebraic_constraints_BRST_extended} reads
\begin{equation}
\label{helicity_algebraic_constraints_light-cone}
\begin{gathered} 
  a_i^m \frac{\partial}{\partial a_j^m} \psi = 0\;, \qquad 1 \le i < j \le n\;,
  \qquad
  \fo^m \frac{\partial}{\partial a_i^m} \psi = 0\;, \qquad i = 1, \ldots, n\;,\\
  a_i^m \frac{\partial}{\partial a_i^m} \psi = s_i \psi\;, \qquad i = 1, \ldots, n \;.
\end{gathered}
\end{equation}
Thus, the field content is given by  spinor-tensors with transversal components only, and subject to the light-cone condition  $p^2 = 0$ and the algebraic conditions \eqref{helicity_algebraic_constraints_light-cone}.

The $\Liealg{iso}(d-1,1)$ generators in the light-cone basis are split into two groups of kinematical $G_\text{kin} = (P^+, P^m, M^{+m}, M^{+-}, M^{mk})$ and dynamical $G_\text{dyn} = (P^-, M^{-k})$ generators. After  reduction  to the $\Omega_{-1}$-cohomology both types of generators give rise to the reduced generators $\widetilde G_\text{kin}$ and $\widetilde G_\text{dyn}$ defined on the subspace $\cE$. While $\widetilde G_\text{kin}$ retain its form upon the reduction, the explicit expressions for $\widetilde G_\text{dyn}$ are given by
\begin{equation}
\label{dyn}
\widetilde P^- = -	\frac{p_m p^m}{2 p^+}\;,
\qquad \widetilde M^{-m}
  = \left( \frac12 - \frac{\partial}{\partial p^+} \right) p^m + \frac{\partial}{\partial p_m} \frac{p_k p^k}{2 p^+}
  + \frac{1}{p^+} \left( S^{mk} p_k \right)\;,
\end{equation}
where the elements 
\begin{equation}
  S^{mn} = a^m \frac{\partial}{\partial a_n} + \frac14 \fo^m \fo^n - \left( m \leftrightarrow n \right)\;,
\end{equation}
form a little Wigner algebra $\Liealg{o}(d-2)$ with the standard commutation relations.

\section{Continuous spin fermionic fields}
\label{sec:Fcon}

Now, we  turn to the continuous spin fermionic field system introduced in Section \bref{sec:short}. Based on our analysis of the standard fermionic fields in Section \bref{sec:FHF}  we propose the deformed triplet formulation and describe its metric-like and light-cone reductions.

\subsection{Deformed triplet formulation }

As before, there are  anticommuting ghost variables $c_0,c_i,b_i$ and commuting ghost variable $\alpha$, with ghosts numbers $\gh{c_0} = \gh{c_i} = \gh{\alpha} = 1$, $\gh{b_i} = -1$.
The BRST operator associated to the constraint system \eqref{Dirac}, \eqref{gammatr}, and \eqref{gauge} is given by  
\begin{equation} 
\label{continuous_BRST_full}
  \brst = \alpha \Dirac + c_0 \Box + c_i D^i + (D^\dag_i + \mu_i) \frac{\partial}{\partial b_i} - \alpha \alpha \frac{\partial}{\partial c_0} - c_i \frac{\partial}{\partial b_i} \frac{\partial}{\partial c_0}\;.
\end{equation}
It acts on the subspace singled out by the ghost extended algebraic constraints \eqref{youngs}
\be
\begin{aligned} 
\label{continuous_algebraic_constraints_BRST_extended}
&\left( \Upsilon^i + \nu^i \ao - 2 \alpha \frac{\partial}{\partial c_i} + \frac{\partial}{\partial \alpha} \frac{\partial}{\partial b_i} \right) \Psi = 0, \qquad i = 1, \ldots, n\;,\\
&\left( N_m + b_m \frac{\partial}{\partial b_m} + c_m \frac{\partial}{\partial c_m} \right) \Psi = s_m \Psi, \qquad m = 2, \ldots, n\;,\\
&\left( N_m{}^k + b_m \frac{\partial}{\partial b_k} + c_m \frac{\partial}{\partial c_k} \right) \Psi = 0\;,  \qquad m,k = 2, \ldots, n \quad (m < k)\;.
\end{aligned}
\ee
Recall that $\mu_i = \delta_{i1}\mu$ and $\nu^i =\delta^{i1}\nu$ so that the deformed triplet operator \eqref{continuous_BRST_full} differs from the undeformed triplet operator \eqref{helicity_BRST_full} only by the term $\mu \frac{\partial }{\partial b_1}$.  Also, just as in the helicity case,   \eqref{continuous_BRST_full} retains the property of being nilpotent on the whole space, not only on the subspace  \eqref{continuous_algebraic_constraints_BRST_extended}.

\subsection{Metric-like formulation}

Similarly to the helicity case of Section \bref{sec:redu} the representation space can be endowed with an additional grading with respect to the ghost $c_0$ so that the lowest component $\brst_{-1}$ of the deformed BRST operator \eqref{continuous_BRST_full} remains the same, while the deformation term enters $\brst_{0}$. Furthermore, one can check that $H(\brst_{-1})$ cohomology remains unchanged except that now the entire subspace is singled out by the deformed constraints \eqref{continuous_algebraic_constraints_BRST_extended} rather than the undeformed ones.

Repeating the same steps as in the helicity case we obtain  the reduced equations of motion
\begin{equation} 
\label{continuous_metric-like}
  \left[ \Dirac - (D^\dag_i + \mu_i) (\Upsilon^i + \nu^i \ao) \right] \psi = 0\;,
\end{equation}
which are invariant with respect to the gauge transformations
\begin{equation}
\label{gauge_con}
\delta \psi = (D^\dag_i + \mu_i) \epsilon^i\;,
\end{equation}
where both fields and parameters  are subjected to the modified trace conditions
\begin{equation}
\label{def_con}
  \mathbf \Upsilon^{(i} \mathbf \Upsilon^{j} \mathbf \Upsilon^{k)} \psi = 0\,,\qquad
  \mathbf \Upsilon^{(i} \epsilon^{j)} = 0\,,
\end{equation}
where the notation for the deformed gamma-trace operator $\mathbf \Upsilon^i = \Upsilon^i + \nu^i \ao$ is introduced. In the case of spin-$\frac{1}{2}$ continuous spin field (i.e. $n=1$ so that there is only one commuting auxiliary variable $a^m_1$) the equation \eqref{continuous_metric-like} reproduces the field equations obtained in \cite{Bekaert:2005in,Najafizadeh:2015uxa}.

It is worth noting that the equation \eqref{continuous_metric-like} can be squared  to yield 
\begin{equation}
  \left[ \Box - (D^\dag_i + \mu_i) D^i + \frac12 (D^\dag_i + \mu_i) (D^\dag_j + \mu_j) (T^{ij} + \nu^i \nu^j) \right] \psi = 0\;,
\end{equation}
which is the bosonic continuous spin metric-like equation \cite{Alkalaev:2017hvj}.

To conclude this section we describe the field space in the case of spin-$\frac{1}{2}$ continuous spin field. The corresponding metric-like fields were previously considered in \cite{Metsaev:2017ytk,Khabarov:2017lth}. Solving the deformed constraints \eqref{def_con} we find that both the fields and parameters can be equivalently represented as infinite chains of the Fang-Fronsdal tensors
\be
\label{covdec}
\begin{aligned}
& \psi \;: = \; \bigoplus_{k=0}^\infty \psi_{(k)}\;,\qquad \Upsilon^3 \psi_{(k)} = 0\;,\\
& \epsilon \;: = \; \bigoplus_{k=0}^\infty \epsilon_{(k)}\;, \;\;\qquad \Upsilon \epsilon_{(k)} = 0\;.
\end{aligned}
\ee   
Technically, the above decompositions are obtained by substituting \eqref{continuous_gamma-trace_decomposition} into \eqref{def_con} and solving the recurrence  equations for expansion coefficients. Using \eqref{covdec} in the metric-like equations \eqref{continuous_metric-like} and the gauge transformations \eqref{gauge_con} gives rise to the Schuster-Toro type equations invariant with respect to the $\mu$-deformed gauge transformations  \cite{Metsaev:2017ytk}.

\subsection{Light-cone formulation}

We start with the BRST operator \eqref{continuous_BRST_full} and use the grading \eqref{helicity_light-cone_grading}. In the considered functional class \eqref{continuous_gamma-trace_decomposition} any element has a finite grading because $\deg \Upsilon_i = 0$, and, by assumption, the degree of coefficients in \eqref{continuous_gamma-trace_decomposition} is bounded so that we can use the homological reduction technique.

The operator \eqref{continuous_BRST_full} decomposes into the homogeneous degree components as $\brst = \brst_{-1} + \brst_0 + \brst_1 + \brst_2 + \brst_3$, where
\begin{equation}
\begin{gathered}
  \brst_{-1} = p^+ \left( 2 \alpha \frac{\partial}{\partial \fo^+} + c_i \frac{\partial}{\partial a_i^+} + a_i^- \frac{\partial}{\partial b_i} \right)\,,\qquad
  \brst_0 = c_0 \Box\,,\\
  \brst_1 = \alpha \fo^m p_m + c_i p^m \frac{\partial}{\partial a_i^m} + (a_i^m p_m + \mu_i) \frac{\partial}{\partial b_i}\,,\qquad
  \brst_2 = - \left( \alpha \alpha + c_i \frac{\partial}{\partial b_i} \right) \frac{\partial}{\partial c_0}\,,\\
  \brst_3 = p^- \left( \alpha \fo^+ + c_i \frac{\partial}{\partial a_i^-} + a_i^+ \frac{\partial}{\partial b_i} \right)\,.
\end{gathered}
\end{equation}
The deformation term $\mu\frac{\partial}{\partial b_1}$ is contained in $\brst_1$ only, and, therefore, the reduced BRST operator is the same as in the helicity case, 
\begin{equation}
  \widetilde \Omega = c_0 \Box\;,
\end{equation}
The light-cone algebraic constraint following from \eqref{continuous_algebraic_constraints_BRST_extended} read
\be
\begin{aligned} 
\label{continuous_algebraic_constraints_light-cone}
&\hspace{-4mm}\left( \fo^m \frac{\partial}{\partial a_i^m} + \nu^i \Gamma \right) \psi = 0\,, \qquad i = 1, \ldots, n\;,\\
& a_i^m \frac{\partial}{\partial a_j^m} \psi = 0\,, \qquad 2 \le i < j \le n\;,\\
& a_i^m \frac{\partial}{\partial a_i^m} \psi = s_i \psi\,, \qquad i = 2, \ldots, n\;.
\end{aligned}
\ee
Thus, the light-cone fields $\psi$ are $\Liealg{o}(d-2)$ spinor-tensors (with doubled spectrum for the odd $d$) subjected to the light-cone condition  $p^2 = 0$ and the algebraic conditions \eqref{continuous_algebraic_constraints_light-cone}.

\paragraph{Poincar\'e algebra realization.} Just as in Section \bref{sec:light_helicity} we reduce the Poincar\'e generators to obtain the same expressions for all Poincar\'e generators except for the dynamical generators, cf. \eqref{dyn}. The translation generator $\widetilde P^-$ remains the same in the continuous spin case, while the Lorentz generator $\widetilde M^{-m}$ gets a new contribution proportional to the deformation parameter $\mu$,   
\begin{equation}
  \widetilde M^{-m}
  = \left( \frac12 - \frac{\partial}{\partial p^+} \right) p^m + \frac{\partial}{\partial p_m} \frac{p_k p^k}{2 p^+}
  + \frac{1}{p^+} \left( S^{mk} p_k - H^m \right),
\end{equation}
where
\begin{equation}
  S^{mn} = a^m_i \frac{\partial}{\partial a_{in}} + \frac14 \fo^m \fo^n - \left( m \leftrightarrow n \right),\qquad
  H^m = \mu \frac{\partial}{\partial a_{1 m}}\;.
\end{equation}
Elements $S^{mn}$ and $H^m$ satisfy the commutation relations
\be
\label{iso}
[S^{kl}, S^{ps}] = \delta^{kp}S^{ls}+ \text{3 terms}\;, 
\qquad
[S^{kl}, H^p] = \delta^{kp}H^l -\delta^{lp}H^k\;, 
\qquad
[H^k, H^l] = 0\;,
\ee
thereby forming the $\Liealg{iso}(d-2)$ algebra. 

Let us evaluate the first two Casimir operators of the $\Liealg{iso}(d-2)$ algebra on the subspace \eqref{continuous_algebraic_constraints_light-cone} following the analogous considerations in Section~\bref{sec:short}. We find 
\begin{multline}
\label{Casimirs_light-cone_evaluated}
c_2 \equiv  H^2 \approx - \mu^2 \nu^2\,,\qquad \qquad
c_4 \equiv (HS)^2 - \frac12 H^2 S^2 \approx
\\ \approx- \mu^2 \nu^2
\left(
  \sum_{i=2}^n s_i (s_i + d - 1 - 2i)
+ \left\{ \sum_{i=2}^n s_i + \frac{(d - 3) (d - 4)}{8} \right\}
\right)\,,
\end{multline}
where $H^2 = H^m H_m$, $S^2 = S_{mk}S^{mk}$,  $(HS)^m = H_k S^{km}$. See \eqref{Casimir2_light-cone} and \eqref{Casimir4_light-cone} for explicit expressions for the Casimir operators $c_2$ and $c_4$ in terms of the $\Liealg{osp}(1|2n)$ basis elements. Higher order Casimir operators can be found analogously. Note that if we drop terms in the curly brackets we get the Casimir eigenvalue for a bosonic field.

Let us solve the modified trace constraint explicitly and describe the $\Liealg{o}(d-2)$ structure of the field space. For simplicity, we consider the case of spin-$\frac{1}{2}$ continuous spin field, where all the $\Liealg{o}(d-3)$ spin weights are zero. Solving the constraints \eqref{continuous_algebraic_constraints_light-cone} we find that the original light-cone spinor-tensor can be represented as an infinite direct sum 
\be
\label{lc_space}
\psi\;\coloneqq\;   \bigoplus_{k=0}^\infty \psi_{(k)}\;,
\ee  
where $\psi_{(k)}$ are $\Liealg{o}(d-2)$ totally symmetric rank-$k$ spinor-tensors satisfying the undeformed gamma-tracelessness condition. Both the $\Liealg{o}(d-2)$ light-cone and $\Liealg{o}(d-1,1)$ covariantized form of the above infinite-dimensional space appeared in the earlier literature \cite{Brink:2002zx,Metsaev:2017ytk,Khabarov:2017lth}, cf. \eqref{covdec}. For  higher values of $n$ which correspond to non-zero spins the modified  constraints \eqref{continuous_algebraic_constraints_light-cone} can be solved to obtain extended infinite-dimensional field spaces along the lines of \cite{Alkalaev:2017hvj}.     

\section{Weyl and gauge modules}\label{sec:modules}
\label{sec:module}

A linear gauge system is essentially determined by the space of gauge inequivalent formal solutions to the equations of motion, known as Weyl module, and the space of (higher-order) global reducibility parameters, known as  gauge module. These spaces are usually considered as modules over the space-time global symmetry algebra.  In particular, if the gauge module vanishes the system is non-gauge, i.e. all the gauge symmetries are  Stueckelberg-like. Note also, that if the gauge module vanishes and the space-time global symmetries (e.g. Poincar\'e or AdS or conformal) act transitively, the system is entirely determined by the Weyl module structure. This property is manifest in the unfolded approach.\footnote{For a  review of the unfolded approach, see e.g.~\cite{Bekaert:2005vh}. Within the present framework, more details on the gauge and Weyl modules can be found in~\cite{Barnich:2015tma,Chekmenev:2015kzf} and references therein.} 

We are now interested in the gauge and Weyl modules of the fermionic (continuous) spin system. To this end, we extend the analysis of~\cite{Alkalaev:2008gi,Alkalaev:2017hvj} to the case of fermionic fields. To study formal solutions in this section we replace space-time coordinates $x^a$ by formal coordinates $y^a$. In particular, it is implicitly assumed that in all the expressions for fields, parameters, operators, etc. $x^a$ and $\dl{x^a}$ are replaced with $y^a$ and $\dl{y^a}$, respectively. Moreover, instead of smooth functions in $x^a$ we work with formal power series in $y^a$. The relevant space is that of formal series in $y^a$ and $a^a_i$ with coefficients in $\cS$ such that for a given element the coefficients of the trace decomposition are polynomials in $a^a_i$.

The gauge and Weyl modules can be defined as the cohomology $H^k(Q,\cH_0)$ of the continuous spin generalization 
\begin{equation}
\label{fiveQ}
Q = \left(\sd_i+\mu_i\right) \dl{b_i}\;, \qquad \text{where}\qquad \sd_i = a^i_a\dl{y^a}\;,
\qquad  i = 1,...,n\;,
\end{equation} 
of the BRST operator~\eqref{helicity_BRST}.  It is defined on the subspace $\cH_0$ singled out by the Dirac constraint  \eqref{Dirac}, modified trace constraint \eqref{gammatr}, and BRST extended Young symmetry and spin weight constraints
\begin{equation}
\label{def-tr}
\Dirac \psi=0\,, \quad (\Upsilon^i+\nu^i\Gamma)\psi=0\,, \quad D^i\psi=0\,,\quad \cN_m{}^k\psi=0\,,\quad (\cN_m-s_m)\psi=0\,,
\end{equation} 
where $i = 1,\ldots,n$, and $m,k = 2,\ldots,n$ for continuous spin case ($\mu,\nu \neq 0$), and $m,k = 1,\ldots,n$ for the helicity one ($\mu,\nu=0$).  The Weyl module is the zero ghost number cohomology $H^0(Q,\cH_0)$, the gauge module is a collection of modules identified with negative ghost degree cohomology $H^k(Q,\cH_0)$ at $k<0$ \cite{Barnich:2004cr,Alkalaev:2008gi,Alkalaev:2009vm,Alkalaev:2011zv}.

To compute the $Q$-cohomology we realize the space $\cH_0$ defined by~\eqref{def-tr} as a subspace of the tensor product 
\begin{equation}
\begin{gathered}
\label{def-ptr}
\cH=\cS \otimes \cG\,, \\
\cG=\{\psi: (T^{ij}+\nu^i\nu^j)\psi=D^i\psi=\Box\psi=N_m{}^k\psi=(N_m-s_m)\psi=0
\}\;,
\end{gathered}
\end{equation} 
where $\psi$ is a formal series in $a^a_i$ and $y^a$ with coefficients in $\mathbb{C}$  such that the coefficients of its trace decomposition are polynomials in $a^a_i$. Indeed, $\cH_0$ is just  a subspace of~\eqref{def-ptr} singled out by
\begin{equation}
 (\Upsilon^i + \nu^i \ao)\psi=\Dirac\psi=0\,.
\end{equation} 
As a next step, we note that it is enough to compute the $Q$-cohomology in $\cH$. Indeed, $\cH$ can be represented as a direct sum of $\cH_0$ and the complementary subspace $\cH_1$ in such a way that $Q$ preserves both subspaces. As $\cH_1$ one can take a subspace of elements that can be represented as $(\gamma \cdot a) \alpha+(\gamma\cdot y)\beta$ for some $\alpha, \beta\in \cH_0$
(the dot denotes summation over Lorentz indices). As representatives of the $Q$-cohomology in $\cH_0$ one can take those representatives of the $Q$-cohomology in $\cH$ that belong to $\cH_0$.

Finally, the action of $Q$ on $\cS \otimes \cG$  originates from the action of $Q$ on $\cG$ because $Q$ does not affect $\cS$, and, hence, the $Q$-cohomology in $\cH$ is just a tensor product of $\cS$ with the 
$Q$-cohomology in $\cG$. In its turn, the  $Q$-cohomology in $\cG$ is known for both helicity fields~\cite{Alkalaev:2008gi} and continuous spin fields~\cite{Alkalaev:2017hvj}.

\subsection{$Q$-cohomology for helicity fermionic fields}

For $\mu,\nu = 0$, the  space $\cG$ defined by~\eqref{def-ptr} is precisely the representation space involved in describing bosonic helicity fields  of general Young symmetry type. Let us spell out the explicit description~\cite{Alkalaev:2008gi} of the representatives of $Q$-cohomology classes. Introduce the following subspaces $\cM_{k}\subset \cG$, $k=0,\ldots,n-1$
\begin{equation}
\begin{gathered}
 \cM_k=\{\psi\in \cG\,:\, \gh{\psi}=-k\,, \quad (y\cdot\dl{a_m})\psi=0\,, \quad  \sd_l \psi=0\}\,, \\
 \qquad m=1,\ldots, n-k-1\,,\quad l=n-k,\ldots,n-1 \,.
\end{gathered}
\end{equation}
For any $k=1,\ldots, n-1$ each cohomology class from $H^{-k}(Q,\cG)$ has a unique representative belonging to $\cM_k$. 

Taking into account the above characterization of the cohomology classes one concludes that 
\begin{equation}
 H^{-k}(Q,\cH_0)\simeq \{\psi\in \cM_k\otimes \cS\,:\, \Upsilon^i\psi=\Dirac\psi=0\}=(\cM_k\otimes \cS) \cap \cH_0\,.
\end{equation} 
Note that the last equality makes sense as $\cM_k$ is naturally a subspace in $\cG$.

\subsection{$Q$-cohomology for continuous spin fields}

In the case $\mu,\nu \neq 0$ the subspace $\cG$ is again a relevant subspace. The cohomology $H^k(Q,\cG)$ was studied in~\cite{Alkalaev:2017hvj}, where it was shown
that $H^k(Q,\cG)=0$ for $k<0$ and $H^0(Q,\cG)\neq 0$ with our choice of the functional class. Since
\begin{equation}
H^k(Q,\cH_0)= (H^k(Q,\cG)\otimes \cS)\cap \cH_0\;,
\end{equation} 
it  follows that $H^k(Q,\cH)=0$ at $k<0$. In particular, we conclude that fermionic continuous spin fields are  not gauge fields as well, i.e. all gauge fields present in the formulation are actually Stueckelberg ones. 

The Weyl module for  bosonic continuous spin fields is given by $H^0(Q,\cG)\neq 0$. Let us show that the same applies to $H^0(Q,\cH_0)$. Let $\psi$ be a nontrivial representative of $H^0(Q,\cG)$, then one can choose $\xi\in\cS$ such that $\psi_0=(\psi \otimes \xi)\cap \cH_0$ is also non-vanishing. Furthermore, $Q\psi_0=0$ because $Q$ preserves both $\cH_0$ and $\cH_1$.

\vspace{5mm}
\noindent \textbf{Acknowledgements.} We are grateful to R. Metsaev, M. Najafizadeh, A. Reshetnyak for useful comments. The work of K.A. was supported by the grant RFBR No 17-02-00317. The work of A.C. and M.G. was supported by the grant RFBR No 18-02-01024.

\appendix

\section{$\Liealg{osp}(1|2n)$ commutation relations}
\label{sec:appA}

The basis $\Liealg{osp}(1|2n)$ elements are defined in \eqref{SPgenerators} and \eqref{Fgenerators}. Their non-zero commutation relations  in the even sector are
\be
\label{Brela}
\ba{c}
\dps
[T_I{}^J, T_K{}^L]= \delta_K^J T_I{}^L-\delta_I^LT_K{}^J,
\quad
[T^{IJ}, T_{KL}] = \delta^I_K T_L{}^J+\delta^I_L T_K{}^J+\delta^J_K T_L{}^I+\delta^J_L T_K{}^I,
\\
\\
\dps
[T_K{}^L, T_{IJ}]= \delta_J^L T_{KI}+\delta^L_IT_{KJ} \;,
\quad
[T^{IJ}, T_K{}^L]=\delta^I_K T^{JL}+ \delta^J_KT^{IL} \;,
\ea
\ee
in the odd sector are
\begin{equation} \label{Frela}
  \{\Upsilon_I, \Upsilon_J\} = 2\, T_{IJ}\;,
  \qquad
  \{\Upsilon_I, \Upsilon^J\} = 2\, T_{I}{}^{J}\;,
  \qquad
  \{\Upsilon^I, \Upsilon^J\} = 2\, T^{IJ}\;,
\end{equation}
in the cross-sector are
\begin{equation} \label{Crela}
\begin{array}{c}
  [T_{IJ}, \Upsilon^K] = -\delta_I^K \Upsilon_J  - \delta_J^K \Upsilon_K\;,
  \qquad
  [T^{IJ}, \Upsilon^K] = 0\;,
  \qquad
  [T_I{}^J, \Upsilon^K] = -\delta_I^K \Upsilon_J\;,
  \\
  \\
  
  [T^{IJ}, \Upsilon_K] = \delta_I^K \Upsilon^J + \delta_K^J \Upsilon^I\;,
  \qquad
  [T_{IJ}, \Upsilon_K] = 0\;,
  \qquad
  [T_I{}^J, \Upsilon_K] = \delta_K^J \Upsilon_I\;.
\end{array}
\end{equation}

\section{Casimir operators} 
\label{Casimir operators section}

The quadratic and quartic Casimir operators of the $\Liealg{iso}(p, q)$ algebra are
\begin{equation} \label{Casimirs}
C_2 \Big( \mathfrak{iso}(p, q) \Big) = P_a P^a \equiv P^2\,,
\qquad
C_4 \Big( \mathfrak{iso}(p, q) \Big) = M_{ab} P^b M^{ac} P_c - \frac12 M^2 P^2\,,
\end{equation}
where $P_a$ stands for translation and $M_{ab}$ for rotation generators, respectively. In what follows, we express \eqref{Casimirs} in terms of the $\Liealg{osp}$ basis elements.

\paragraph{Regular spinor-tensor representation.}
Let $\Liealg{iso}(d-1,1)$ basis elements $P_a\,, M_{ab}\,, a,b = 0, ..., d-1$ act as
\begin{equation}
  P_a = \partial_a\;, \quad\qquad 
  M_{ab}
  = x_a \partial_b - x_b \partial_a
  + a_i{}_a \bar a^{i}_b - a_i{}_b \bar a^{i}_a
  + \frac14 (\fo_a \fo_b - \fo_b \fo_a)\,,
\end{equation}
$i = 1, \ldots, n$. Expressing them in terms of the $\Liealg{osp}(1|2n+2)$ basis elements we find 
\begin{equation} \label{Casimir2_regular}
  C_2 \Big( \Liealg{iso}(d-1,1) \Big) = \Box\;,
\end{equation}
\begin{multline} \label{Casimir4_regular}
  C_4 \Big( \Liealg{iso}(d-1,1) \Big)
  =
    \left( (d - n - 2) N_i{}^i + N_j{}^i N_i{}^j - T_{i j} T^{i j} \right) \Box\\
  + T_{i j} D^i D^j + (2 - d) D^\dag_i D^i - 2 D^\dag_j N_i{}^j D^i + D^\dag_i D^\dag_j T^{i j}\\
+ \left\{
    \left( \Upsilon_i D^i - D^\dag_i \Upsilon^i \right) \Dirac
  + \left( N_i{}^i - \Upsilon_i \Upsilon^i + \frac{(d - 1) (d - 2)}{8} \right) \Box
  \right\}\,.
\end{multline}
Note that dropping terms in curly brackets we get the bosonic Casimir operator. Also, the above $\Liealg{osp}(1|2n+2)$ representation holds for any $\Liealg{iso}(k,l)$ with $k+l = d$. 

\paragraph{Light-cone realization.} Let $\Liealg{iso}(d-2)$ basis elements $H_m\,, S_{m k}\,,\; m, k = 0, \ldots, d - 3$ act as
\begin{equation}
	H_m = \mu \frac{\partial}{\partial a_1^m}\,,\qquad\qquad
  S_{mn} = a_{i m} \frac{\partial}{\partial a_i^n} + \frac14 \fo_m \fo_n - \left( m \leftrightarrow n \right)\,,
\end{equation}
$i = 1, \ldots, n$. Then, we find 
\begin{equation} \label{Casimir2_light-cone}
	C_2 \Big( \Liealg{iso}(d - 2) \Big) = \mu^2 T^{11}\,,
\end{equation}
\begin{multline} \label{Casimir4_light-cone}
  C_4 \Big( \Liealg{iso}(d - 2) \Big)
= \mu^2
\bigg(
  \left( (d - n - 2) N_i{}^i + N_j{}^i N_i{}^j - T_{i j} T^{i j} \right) T^{1 1}\\
  + T_{i j} T^{i 1} T^{j 1}
  + (3 - d) N_i{}^1 T^{i 1}
  - 2 N_j{}^1 N_i{}^j T^{i 1}
  + N_i{}^1 N_j{}^1 T^{i j}\\
+ \left\{
    \left( \Upsilon_i T^{i 1} - N_i{}^1 \Upsilon^i \right) \Upsilon^1
  + \left( N_i{}^i - \Upsilon_i \Upsilon^i + \frac{(d - 3) (d - 4)}{8} \right) T^{1 1}
  \right\}
\bigg)\,.
\end{multline}
The  terms in curly brackets are again the fermionic addition to the bosonic Casimir operator.
Expression \eqref{Casimir4_light-cone} goes to \eqref{Casimir4_regular} under substitution
\begin{equation}
\begin{gathered}
	(d - 2) \mapsto d\,,\quad
  (n + 1) \mapsto n\,,\quad
  a_1^m \mapsto x^a\,,\quad
  \mu \frac{\partial}{\partial a_1^m} \mapsto \frac{\partial}{\partial x^a}\,,\\
  a_2^m \mapsto a_1^a\,,\quad
  \frac{\partial}{\partial a_2^m} \mapsto \frac{\partial}{\partial a_1^a}\,,\quad
  \ldots\,,\quad
  a_{n+1}^m \mapsto a_n^a\,,\quad
  \frac{\partial}{\partial a_{n+1}^m} \mapsto \frac{\partial}{\partial a_n^a}\,.
\end{gathered}
\end{equation}

%\bibliographystyle{JHEP}
%\bibliography{HSmaster}

\providecommand{\href}[2]{#2}\begingroup\raggedright\endgroup

\end{document}